\documentclass[twocolumn,superscriptaddress,nofootinbib,floatfix,raggedbottom]{revtex4-2}

\usepackage{amsmath,amssymb,bm}
\usepackage{graphicx}
\usepackage{booktabs}
\usepackage{array}[=2016-10-06]
\usepackage{needspace}
\usepackage{url}
\usepackage{microtype}
\usepackage[colorlinks=true,linkcolor=blue,citecolor=blue,urlcolor=blue]{hyperref}

\graphicspath{{./}}
\renewcommand{\arraystretch}{1.12}

\makeatletter
\renewcommand{\subsection}{
  \@startsection{subsection}{2}{\z@}
  {16pt}{8pt}
  {\normalfont\small\bfseries\centering}}
\makeatother

\newcommand{\vect}[1]{\bm{#1}}
\newcommand{\dd}{\mathrm{d}}

\newcommand{\epszero}{\epsilon_0}
\newcommand{\warpx}{WarpX}
\newcommand{\pmsl}{\texttt{Algoplasma\allowbreak -pic\allowbreak -3d\allowbreak -raz}}
\newcommand{\figref}[1]{\hyperref[#1]{Fig.~\ref*{#1}}}
\newcommand{\tabref}[1]{\hyperref[#1]{Table~\ref*{#1}}}

\begin{document}

\title{Annular-curvature effects on high-frequency azimuthal fluctuations in 3D PIC simulations of Hall-thruster plasmas}

\author{Yingjie Chen}
\affiliation{School of Energy Science and Engineering, Harbin Institute of Technology, Harbin 150001, People's Republic of China}

\author{Yinjian Zhao}
\email[Corresponding author: ]{zhaoyinjian@hit.edu.cn} 
\affiliation{School of Energy Science and Engineering, Harbin Institute of Technology, Harbin 150001, People's Republic of China}

\begin{abstract}
Azimuthal instabilities and associated plasma fluctuations play an important role in anomalous cross-field electron transport in Hall thrusters. Kinetic studies often approximate the annular discharge using local Cartesian domains, but the influence of the neglected curvature on three-dimensional fluctuation dynamics remains insufficiently quantified.\ 
Here, three-dimensional cylindrical electrostatic particle-in-cell simulations assess this geometric effect in a reduced channel--plume model by varying the minimum computational-domain radius $r_{\min}$ at fixed channel width and imposed magnetic-field profile.\ 
The dominant channel wavenumbers on the centerline-reference scale remain similar, while pronounced radial amplitude asymmetry develops in the smaller-radius configurations. At $r_{\min}=10~\mathrm{cm}$ and $2~\mathrm{cm}$, the outer-channel electric-field amplitudes in the dominant wavenumber band exceed the inner-channel values by $26\%$ and $53\%$, respectively. In the plume, low-wavenumber fluctuations favor the inner region, whereas high-wavenumber components favor the outer region. Density--electric-field covariance is preferentially concentrated in the outer channel and inner plume, with outer-to-inner magnitude ratios of 2.71 and 0.58 at $r_{\min}=2~\mathrm{cm}$. These persistent spatial preferences connect the geometric redistribution of fluctuations to the correlated axial $E\times B$ drift contribution.\ 
For the $1.28~\mathrm{cm}$ channel width studied, these results support retaining cylindrical geometry at $r_{\min}=10~\mathrm{cm}$ and $2~\mathrm{cm}$ to resolve local fluctuations and transport-related correlations, even when the imposed magnetic-field profile is unchanged.
\end{abstract}


\maketitle
\raggedbottom
\setlength{\parskip}{0pt}
\clubpenalty=10000
\widowpenalty=10000
\displaywidowpenalty=10000
\setlength{\abovedisplayskip}{7pt plus 1pt minus 2pt}
\setlength{\belowdisplayskip}{7pt plus 1pt minus 2pt}
\setlength{\abovedisplayshortskip}{3pt plus 1pt}
\setlength{\belowdisplayshortskip}{5pt plus 1pt minus 1pt}

\section{Introduction}

Anomalous cross-field electron transport remains a central difficulty in predictive Hall-thruster modeling because the effective transport often exceeds classical collisional estimates
\cite{haraOverviewDischargePlasma2019,Kaganovich2020Perspective,Ahedo2023Review,Taccogna2023Perspective}.
Theoretical descriptions of turbulence-enhanced electron mobility
\cite{lafleurTheoryAnomalousElectron2016,lafleurTheoryAnomalousElectron2016a}
and wave--particle momentum exchange \cite{Mikellides2024Momentum} identify plasma
fluctuations as an important contributor. In a predominantly radial magnetic field,
an azimuthal electric-field fluctuation produces an axial \(E\times B\) drift.
Correlations between this drift and electron-density fluctuations can therefore
contribute to a net axial electron flux \cite{Lafleur2017PSST,Croes2017PSST}.
This contribution depends on fluctuation amplitudes and relative phases, while its
distribution across the discharge depends on where these correlations develop.
Annular curvature may modify both the background plasma distribution and the spatial development of fluctuations, making its influence relevant to modeling electron transport.

Assessing this geometric influence requires resolving the nonlinear evolution and
spatial structure of the fluctuations. The electron drift instability (EDI), also
commonly termed the electron cyclotron drift instability (ECDI), is of particular
interest in this context \cite{Boeuf2018POP,Charoy2019PSST,Villafana2021PSST}.
Recent experiments have measured its growth and nonlinear energy transfer,
identifying three-wave interactions and an inverse cascade toward lower frequencies
\cite{Brown2023PRL}. Measurements of multidimensional wave-vector structure also
show that high-frequency fluctuations are not confined to purely azimuthal
propagation \cite{Tsikata2010}. Three-dimensional particle-in-cell (PIC)
simulations can retain these coupled dynamics by evolving particle kinetics and
electric fields self-consistently \cite{Zhao2026Review,Villafana2023POP}.
Our previous three-dimensional PIC studies have examined the effects of plume-region arrangement and magnetic-field configuration on azimuthal instabilities
\cite{Chen2025POP,Zhong2026RZ}.
Related studies have identified persistent near-wall transport pathways and characterized spatial transport variations using density--electric-field correlations
\cite{Liu2026NearWall,Zhao2026MagneticTransport}.

At a fixed channel width, decreasing the annular radius increases the relative
variation of circumference across the channel. Cylindrical radial studies have
identified pronounced radial asymmetries arising from geometrical expansion,
magnetic-mirror forces, and centrifugal effects associated with azimuthal electron
motion \cite{DominguezVazquez2018}. Their substantial reduction in a corresponding
planar configuration demonstrates the importance of the cylindrical setting
\cite{DominguezVazquez2019}. Annular radial--azimuthal simulations have further shown that the azimuthal domain
size influences nonlinear mode transitions and anomalous transport, while
centrifugal effects associated with curved electron trajectories modify the
radial plasma structure \cite{senguptaModeTransitionsNonlinear2020}.
More recently, axisymmetric radial--axial simulations varied chamber diameter at
fixed channel width and found stronger radial asymmetries together with changes
in plasma density and wall fluxes at smaller diameters \cite{Petronio2026Curvature}.
That model used a radial magnetic field proportional to \(1/r\) and supplementary
scattering to represent anomalous transport. These studies establish connections
between annular geometry, plasma nonuniformity, and transport, while resolving
different parts of the underlying dynamics.

Taccogna and Minelli explicitly examined cylindrical curvature in a geometrically scaled three-dimensional discharge-channel model by comparing cylindrical and Cartesian formulations
\cite{taccognaThreedimensionalParticleincellModel2018}.
Their curvature assessment used azimuthally averaged radial potential and density profiles at a selected axial location in the acceleration region, where the two formulations yielded similar results.
Although the study also resolved the axial development of azimuthal fluctuations, it did not systematically vary the radius-to-channel-width ratio, and its domain ended at the channel exit. The scaling changed magnetic-field strength and neutral density while preserving the radius-to-width ratio, and thus differed from an independent variation of annular curvature.

A controlled radius scan under a common imposed magnetic-field profile can extend this comparison to the spatial redistribution of fluctuations and density--field correlations across the channel and plume, particularly for compact annular configurations in which the radius is not large relative to the channel width.
The channel walls constrain radial development, whereas the plasma can expand into
a broader radial region beyond the exit. Curvature-induced changes may therefore
vary with axial position. Resolving this dependence is important before detailed
quantitative comparisons with experiments.

Building on our previous three-dimensional studies, we develop the cylindrical
electrostatic PIC code \pmsl{} incorporating selected numerical components from AlgoPlasma
\cite{AlgoPlasma2026}. The cylindrical implementation is first compared with
\warpx{} in the near-Cartesian limit \cite{vayWarpXNewExascale2018} to assess
consistency between the two implementations before interpreting curvature-dependent
changes. We then vary the annular radius while retaining the channel width and a
common imposed magnetic-field profile, allowing radial, azimuthal, and axial
dynamics to evolve self-consistently.
Spatially resolved fluctuation spectra and density--field correlations are compared to assess the range of radius-to-channel-width ratios over which curvature materially affects these quantities under the prescribed conditions, informing geometry choices for subsequent experimental comparisons.
Section~II describes the model and setup, Section~III presents the cross-code and
curvature comparisons, and Section~IV discusses the findings and conclusions.

\section{Numerical model and simulation setup}
\label{sec:method}

\subsection{Physical model and common setup}

Both the Cartesian and cylindrical-coordinate calculations employ the same electrostatic
particle-in-cell (PIC) model. The cylindrical calculations use \pmsl{}.
The electric potential $\phi$ is obtained from Poisson's equation,
\begin{equation}
  \nabla^2 \phi = - \frac{\rho}{\epszero},
  \label{eq:poisson}
\end{equation}
where $\rho$ is the local charge density and $\epszero$ is the vacuum permittivity. The electric
field is then calculated as
\begin{equation}
  \vect{E} = - \nabla \phi .
  \label{eq:electric-field}
\end{equation}
Each macroparticle $p$ is advanced according to
\begin{equation}
  \frac{\dd \vect{x}_p}{\dd t} = \vect{v}_p,
  \label{eq:particle-position}
\end{equation}
\begin{equation}
  \frac{\dd \vect{v}_p}{\dd t}
  =
  \frac{q_p}{m_p}
  \left(
  \vect{E}_p + \vect{v}_p \times \vect{B}_p
  \right),
  \label{eq:particle-motion}
\end{equation}
where $q_p$, $m_p$, $\vect{x}_p$, and $\vect{v}_p$ are the charge, mass, position, and velocity
of particle $p$, respectively, and $\vect{E}_p$ and $\vect{B}_p$ are the electric and magnetic
fields interpolated to the particle position. Linear particle shapes are used for charge
deposition.

The plasma consists of electrons and singly charged xenon ions. The initial plasma density is
$n_0$, and the initial electron and ion temperatures are $T_{e0}$ and $T_{i0}$, respectively.
Electron--ion pairs are generated inside a prescribed pair-source region that stands in for
self-consistent ionization. In the Cartesian reference configuration, the volumetric pair-production rate is uniform in the azimuthal coordinate $y$ and is given by
\begin{equation}
S(x,z)=
\begin{cases}
S_0 C_x(x) C_z(z), & (x,z)\in\Omega_s,\\
0, & \mathrm{otherwise}.
\end{cases}
\label{eq:ionization-source}
\end{equation}
where $S_0$ is fixed by the prescribed total source current rather than treated as an
independent control parameter,
$\Omega_s=[x_1,x_2]\times[z_1,z_2]$ is the radial--axial cross section of the source region. The shape functions are
$C_x(x)=\cos[\pi(x-x_m)/(x_2-x_1)]$ and
$C_z(z)=\cos[\pi(z-z_m)/(z_2-z_1)]$. Here $x_m$ and $z_m$ denote the radial and axial centers
of the source region. The source rectangle spans
$x=7.296$--$18.304~\mathrm{mm}$ and $z=2.5$--$10~\mathrm{mm}$.
In the cylindrical cases, the source is likewise uniform in $\theta$, with $S(r,z)=S_0C_x(x^*)C_z(z)$ over the same local source rectangle and zero outside, where $x^*=r-r_{\min}$. Thus its position relative
to the channel walls and its axial and radial extents are preserved.
Ionization is represented by the prescribed
electron--ion pair source in Eq.~\eqref{eq:ionization-source}.

The prescribed total source current is
\begin{equation}
  I_{\mathrm{tot}} = J_m L_{xc} L_y,
  \label{eq:itot}
\end{equation}
where $J_m$ is the source current density, $L_{xc}$ is the radial channel width, and $L_y$ is
the Cartesian azimuthal periodic length. In the cylindrical sector, $r_c=(r_{\rm ch,in}+r_{\rm ch,out})/2$ is the channel centerline radius, with $r_{\rm ch,in}$ and $r_{\rm ch,out}$ denoting the channel-wall radii.
Then \(L_y=r_c\Delta\theta_{\rm dom}\), and the radial--azimuthal channel area is
\begin{equation}
 \begin{aligned}
 \Delta\theta_{\rm dom}\int_{r_{\rm ch,in}}^{r_{\rm ch,out}}r\,\dd r
 &=r_c\Delta\theta_{\rm dom}(r_{\rm ch,out}-r_{\rm ch,in})\\
 &=L_yL_{xc}.
 \end{aligned}
 \label{eq:cylindrical-source-area}
\end{equation}
Equation~\eqref{eq:itot} therefore specifies the same intended continuous source strength in both coordinate
representations through
\begin{equation}
  \int_{\Omega_{s,3D}}S\,\dd V=\frac{I_{\mathrm{tot}}}{e}.
  \label{eq:source-normalization}
\end{equation}
Here $\Omega_{s,3D}$ denotes the corresponding three-dimensional source region spanning the full periodic azimuthal interval; the physical volume element is $\dd V=\dd x\,\dd y\,\dd z$ in Cartesian coordinates and $\dd V=r\,\dd r\,\dd\theta\,\dd z$ in cylindrical coordinates.
For the bounds in the present calculation this gives
$S_0=9.55\times10^{23}~\mathrm{m^{-3}\,s^{-1}}$. The mean number of electron--ion pairs
generated per time step is consequently
\begin{equation}
  N_{\mathrm{pair}}
  =
  \frac{I_{\mathrm{tot}}\Delta t}{e\,w_0}.
  \label{eq:npair}
\end{equation}
where $\Delta t$ is the time step, $e$ is the elementary charge, and $w_0$ is the
macroparticle weight. In the cylindrical implementation the source-cell selection probability
is proportional to $S(r,z)\,\Delta V$, with
$\Delta V=\tfrac12(r_+^2-r_-^2)\Delta\theta\Delta z$; positions within a selected cell are
sampled uniformly in physical volume. Both the initial particles and newly generated pairs use
zero-drift Maxwellians at $T_e=10~\mathrm{eV}$ and $T_i=0.5~\mathrm{eV}$; the electron and ion
in each generated pair are colocated, while their velocities are sampled independently. The
initial loading is also uniform in physical volume and includes the cylindrical Jacobian. Thus
the value reported in Table~I is a reference mean of 36 particles per active cell for the
uniform near-Cartesian mesh; with the fixed
macroparticle weight, the expected cell count on a non-uniform or appreciably curved mesh is
proportional to the cell's physical volume.

The electrostatic potential is fixed to \(200~\mathrm{V}\) at the anode plane \(z=0\), while
the outlet plane, exposed radial boundaries, and solid channel walls are grounded. The azimuthal
direction is periodic. Electrons and ions striking the anode or solid walls,
or leaving through the open radial or downstream boundaries, are removed.
A separate source at the downstream boundary injects half-Maxwellian electrons
with \(T_e=10~\mathrm{eV}\) and zero drift
using a count-based global feedback condition. At each step the electron and ion macroparticle
populations $N_{e,c}$ and $N_{i,c}$ in the single outlet-adjacent cell layer are summed over all
MPI ranks, and
\begin{equation}
  N_{e,\mathrm{inj}}=\max(0,N_{i,c}-N_{e,c})
  \label{eq:feedback-injection}
\end{equation}
electrons are distributed over the outlet in proportion to physical surface area. These
field and particle conditions are applied in both coordinate representations. The simulations use a collisionless electrostatic model with prescribed pair production and absorbing non-periodic boundaries. Neutral and Coulomb collisions, secondary-electron emission, and dielectric-wall charging are omitted.

The imposed magnetic field is radial and independent of radius and azimuth,
\(\vect B=B_r(z)\vect e_r\), with \(B_z=0\). It is represented by the piecewise Gaussian
\begin{equation}
 B_r(z)=a_s\exp\!\left[-\frac{(z-z_B)^2}{2\sigma_B^2}\right]+b_s,
 \label{eq:magnetic-profile}
\end{equation}
with the branch index
\begin{equation}
 s=\begin{cases}1,&z\le z_B,\\2,&z>z_B.\end{cases}
 \label{eq:magnetic-branch}
\end{equation}
where \(z_B=0.75~\mathrm{cm}\) and \(\sigma_B=0.625~\mathrm{cm}\). The coefficients are fixed by
\(B_r(0)=60~\mathrm{G}\), \(B_r(z_B)=100~\mathrm{G}\), and
\(B_r(2.56~\mathrm{cm})=10~\mathrm{G}\). The fitting point at $z=2.56~\mathrm{cm}$ lies beyond the simulated outlet at $L_z=2.00~\mathrm{cm}$, where $B_r\simeq21~\mathrm{G}$. The same analytic profile, with its maximum at $z=7.5~\mathrm{mm}$ and
$B_r(10~\mathrm{mm})\simeq92.97~\mathrm{G}$, is therefore prescribed across the curvature
scan. 

The purely radial, radius-independent prescription also gives
$\nabla\cdot\mathbf B=B_r/r$ in cylindrical coordinates. The curvature scan uses this idealized radial imposed-field model.
Table~\ref{tab:warpx-parameters} gives the physical inputs and uniform-grid
settings shared by W0, A0, and A0$'$. The listed time step is the electron
step; A0$'$ advances ions every ten electron steps. A1--A4 retain these
physical inputs. Table~\ref{tab:case-list-ap} summarizes their geometry
and mesh changes, as detailed in Sec.~II B.

\begin{table}[!htbp]
\centering
\caption{Physical inputs and uniform-grid settings shared by W0, A0, and A0$'$.
Lengths and cell counts are listed in radial, azimuthal, and axial order;
the electron time step is common, while the ion step ratios are given in
Table~\ref{tab:case-list-ap}.}
\label{tab:warpx-parameters}
\begingroup
\footnotesize
\setlength{\tabcolsep}{2.0pt}
\renewcommand{\arraystretch}{1.05}
\begin{tabular}{>{\raggedright\arraybackslash}p{0.34\columnwidth}
>{\raggedright\arraybackslash}p{0.27\columnwidth}
>{\raggedright\arraybackslash}p{0.29\columnwidth}}
\toprule
Reference parameter & Symbol & W0, A0, A0$'$ value \\
\midrule
Domain size & $L_x\times L_y\times L_z$ & $2.56\times0.64\times2.00~\mathrm{cm^3}$ \\
Grid & $N_x\times N_y\times N_z$ & $256\times64\times200$ \\
Uniform-grid spacing & $\Delta l$ & $0.1~\mathrm{mm}$ \\
Channel size & $L_{xc}\times L_{zc}$ & $1.28\times1.00~\mathrm{cm^2}$ \\
Initial density & $n_0$ & $5.0\times10^{16}~\mathrm{m^{-3}}$ \\
Temperatures & $T_{e0},T_{i0}$ & $10,0.5~\mathrm{eV}$ \\
Applied voltage & $V_a-V_c$ & $200~\mathrm{V}$ \\
Electron time step & $\Delta t$ & $5.0\times10^{-12}~\mathrm{s}$ \\
Particle weight & $w_0$ & $1.389\times10^3$ \\
Initial mean count per active cell & $N_{\mathrm{ppc,ref}}$ & 36/species \\
Source current density & $J_m$ & $400~\mathrm{A\,m^{-2}}$ \\
Magnetic field & $B_r(z)$ & Piecewise Gaussian, Eq.~\eqref{eq:magnetic-profile} \\
\bottomrule
\end{tabular}
\endgroup
\end{table}

\subsection{Coordinate representation and simulation cases}

\begin{table*}[!htbp]
\begingroup
\setlength{\tabcolsep}{2.5pt}
\centering

\caption{Simulation cases, configuration inheritance, and comparison roles.
All A-series cases are computed with
\ \mbox{\pmsl{}}.
All cases use 64 azimuthal cells, with $q=\Delta t_i/\Delta t_e$.
A1--A4 are independently initialized. W0 has no cylindrical curvature parameter.}
\label{tab:case-list-ap}
\footnotesize
\renewcommand{\arraystretch}{1.04}
\begin{tabular}{>{\centering\arraybackslash}p{0.045\textwidth} >{\raggedright\arraybackslash}p{0.14\textwidth}
>{\centering\arraybackslash}p{0.075\textwidth} >{\centering\arraybackslash}p{0.10\textwidth}
>{\raggedright\arraybackslash}p{0.315\textwidth} >{\raggedright\arraybackslash}p{0.265\textwidth}}
\toprule
Case & Formulation & $r_{\min}$ & $\chi$ & Configuration relative to reference & Comparison role \\
\midrule
W0 & Cartesian (\warpx) & -- & -- & Table I baseline; $q=1$ & Cartesian cross-code reference \\
A0 & Cylindrical & $100~\mathrm{m}$ & $6.4\times10^{-5}$ & W0's local dimensions and uniform spacing; $q=1$ & Near-Cartesian counterpart to W0 \\
A0$'$ & Cylindrical & $100~\mathrm{m}$ & $6.4\times10^{-5}$ & A0 geometry and grid; $q=10$; uniform-$\theta$ fast path & Uniform-grid time-integration reference \\
A1 & Cylindrical & $100~\mathrm{m}$ & $6.4\times10^{-5}$ & A0$'$ radius and $q$; non-uniform $r$--$z$ mesh; MPI load balancing & Near-Cartesian production reference, $0$--$15~\mu\mathrm{s}$ \\
A2 & Cylindrical & $1~\mathrm{m}$ & $6.3\times10^{-3}$ & A1 production configuration; reduced radius & Weak curvature, $0$--$15~\mu\mathrm{s}$ \\
A3 & Cylindrical & $10~\mathrm{cm}$ & $5.7\times10^{-2}$ & A1 production configuration; reduced radius & Moderate curvature, $0$--$15~\mu\mathrm{s}$ \\
A4 & Cylindrical & $2~\mathrm{cm}$ & $1.95\times10^{-1}$ & A1 production configuration; reduced radius & Strong curvature, $0$--$15~\mu\mathrm{s}$ \\

\bottomrule
\end{tabular}
\endgroup
\end{table*}

In the WarpX Cartesian reference calculation, the computational coordinates $(x,y,z)$ represent the
radial, azimuthal, and axial directions of the Hall thruster, respectively. The $y$ direction is
periodic. The cylindrical program uses the corresponding physical coordinates $(r,\theta,z)$,
with periodicity in $\theta$. The radial and axial dimensions of the channel and plume are kept
identical in the two representations.

The Cartesian periodic length $L_y$ is used as the reference physical azimuthal length. In the
cylindrical cases, the angular extent is selected at the channel centerline radius $r_c$,
\begin{equation}
  \Delta\theta_{\rm dom}=\frac{L_y}{r_c},
  \label{eq:sector-definition}
\end{equation}
\begin{equation}
  L_{\theta,c}=r_c\Delta\theta_{\rm dom}=L_y.
  \label{eq:sector-length}
\end{equation}
The cylindrical metric is retained in the field solution,
particle advance, charge deposition, and field interpolation. As the absolute radius becomes
large compared with the radial channel width, the metric variation across the domain becomes
negligible and the cylindrical geometry approaches the locally Cartesian limit.

Explicitly, the cylindrical field equations contain the radius-dependent metric terms
\begin{equation}
 \frac{1}{r}\frac{\partial}{\partial r}
 \left(r\frac{\partial\phi}{\partial r}\right)
 +\frac{1}{r^2}\frac{\partial^2\phi}{\partial\theta^2}
 +\frac{\partial^2\phi}{\partial z^2}
 =-\frac{\rho}{\epsilon_0},
 \label{eq:cylindrical-field}
\end{equation}
\begin{equation}
 E_\theta=-\frac{1}{r}\frac{\partial\phi}{\partial\theta}.
 \label{eq:cylindrical-azimuthal-field}
\end{equation}
For a particle of charge $q_p$ and mass $m_p$, the radial and azimuthal velocity equations include
\begin{equation}
 \dot v_r=\frac{q_p}{m_p}(E_r+v_\theta B_z-v_zB_\theta)+\frac{v_\theta^2}{r},
 \label{eq:cylindrical-particle-metric}
\end{equation}
\begin{equation}
 \dot v_\theta=\frac{q_p}{m_p}(E_\theta+v_zB_r-v_rB_z)-\frac{v_rv_\theta}{r}.
 \label{eq:cylindrical-azimuthal-motion}
\end{equation}
The corresponding position equations are $\dot r=v_r$, $\dot\theta=v_\theta/r$, and $\dot z=v_z$. For the prescribed radial field, the axial velocity equation is $\dot v_z=(q_p/m_p)(E_z-v_\theta B_r)$.

The dimensionless annular-curvature parameter is
\begin{equation}
  \chi =
  \frac{r_{\mathrm{ch,out}}-r_{\mathrm{ch,in}}}{2r_c},
  \label{eq:curvature-parameter-general}
\end{equation}
where $r_{\mathrm{ch,in}}$ and $r_{\mathrm{ch,out}}$ are the inner and outer channel radii,
respectively, and
$r_c=(r_{\mathrm{ch,in}}+r_{\mathrm{ch,out}})/2$ is the channel centerline radius. For the
present channel geometry,
\begin{equation}
  r_{\mathrm{ch,in}}=r_{\min}+6.4~\mathrm{mm},
  \label{eq:channel-radii}
\end{equation}
\begin{equation}
  r_{\mathrm{ch,out}}=r_{\min}+19.2~\mathrm{mm},
  \label{eq:channel-outer-radius}
\end{equation}
\begin{equation}
  r_c=r_{\min}+12.8~\mathrm{mm}.
  \label{eq:channel-center-radius}
\end{equation}
and therefore
\begin{equation}
  \chi = \frac{6.4~\mathrm{mm}}{r_{\min}+12.8~\mathrm{mm}}.
  \label{eq:curvature-parameter}
\end{equation}
where $r_{\min}$ is the minimum cylindrical radius of the computational domain. Smaller
$r_{\min}$ therefore corresponds to stronger annular curvature.

Table~\ref{tab:case-list-ap} organizes the numerical configurations as
A0 $\rightarrow$ A0$'$ $\rightarrow$ A1 $\rightarrow$ A2--A4.
A0 is the uniform-grid cylindrical counterpart to W0 at the near-Cartesian
radius $r_{\min}=100~\mathrm{m}$; both advance ions every electron step.
A0$'$ retains A0's geometry and uniform grid, increases the ion-to-electron
step ratio to $q=\Delta t_i/\Delta t_e=10$, and introduces an algebraically
equivalent fast path for the uniform-$\theta$ operations.

A1 retains the near-Cartesian radius and $q=10$ while adopting the
non-uniform production discretization. It keeps
$N_r\times N_\theta\times N_z=256\times64\times200$, with
$\Delta r=0.075$--$0.150~\mathrm{mm}$ and $\Delta z=0.071$--$0.142~\mathrm{mm}$;
refinement is concentrated near the radial channel centerline and
$z\simeq0.69~\mathrm{cm}$. The production implementation uses MPI load balancing.
A2--A4 use the same production treatment and local mesh as A1, but reduce
$r_{\min}$ to $1~\mathrm{m}$, $10~\mathrm{cm}$, and $2~\mathrm{cm}$,
respectively. A0 and A0$'$ use $384$ MPI ranks with two OpenMP threads per
rank; A1--A4 use $128$ ranks with two threads per rank.

This sequence describes configuration inheritance, not continuation from
one case's particle state to another. Each A1--A4 geometry is initialized
independently with the common density, temperature, and initialization seed,
sampling positions in cylindrical physical volume. Each trajectory extends from $t=0$ to
$15~\mu\mathrm{s}$ ($3\,000\,000$ electron steps). Operational continuations use
particle snapshots from the same case and retain the global step and time. Fields are reconstructed from the saved particles; the snapshot format excludes the complete random-number and integrator states. When
assembling a continued history, the accepted continuation replaces the superseded tail of
its predecessor. Temporal variability is evaluated within the common observation window.

\paragraph*{Discrete implementation}
The A-series transfers nodally deposited charge to cell centers for a finite-volume
Poisson solve. Its electric-field components lie on faces normal to their own
directions and at transverse cell centers, with linear gathering in all directions.
W0 filters nodal charge before a nodal embedded-boundary Poisson solve.
Its field components lie at midpoints along their own directions and at transverse
nodes, with zeroth-order gathering along the component direction and linear
gathering transversely.
The cylindrical implementation uses a seven-point finite-volume Poisson operator and the HYPRE StructPFMG solver with a prescribed relative tolerance of $10^{-6}$. Particle advancement applies a Boris rotation in a local Cartesian frame, followed by position and velocity-basis transformations to cylindrical coordinates.
W0 injects cathode electrons before particle advancement, whereas the A-series injects them after advancement and boundary handling. In the A-series, pair creation, charge deposition, the Poisson solve, and field evaluation complete the step. Between ion updates, the stored ion charge is retained and charge from newly created ions is added.

\subsection{Diagnostics}
\label{sec:diagnostics}

The comparison uses three levels of diagnostics. First, the total ion macroparticle number and
corresponding domain-averaged ion density monitor the global source--loss balance and transient
evolution,
\begin{equation}
  \langle n_i\rangle =
  \frac{N_{i,\mathrm{macro}}\,w_i}{V_{\mathrm{act}}},
  \label{eq:domain-averaged-density}
\end{equation}
where $N_{i,\mathrm{macro}}$, $w_i$, and $V_{\mathrm{act}}$ denote the total ion
macroparticle number, ion macroparticle weight, and active plasma volume, respectively.
Second, instantaneous ion density, potential, and azimuthal electric field are compared
on matched physical cuts at $8~\mu\mathrm{s}$ in W0, A0, and A1.  Third, the global
spectral histories retain all seven cases, whereas the quantitative radial spectra
and density--field covariance use only the controlled A1--A4 curvature scan.

For the cylindrical cases, consider one saved field snapshot at physical time
$t$. The radial and axial cell centers are denoted by $r_i$ and $z_\ell$,
where $i$ and $\ell$ are their respective cell indices. At each fixed
$(r_i,z_\ell)$, the azimuthal electric field $E_\theta$ supplies $N_\theta=64$
uniformly spaced samples over the periodic sector of angular width
$\Delta\theta_{\rm dom}$. The sample index is $j=0,\ldots,N_\theta-1$, and
$E_{{\rm azi},j}(r_i,z_\ell,t)=E_\theta(r_i,\theta_j,z_\ell,t)$ denotes the field
at the azimuthal cell center $\theta_j=(j+1/2)\Delta\theta_{\rm dom}/N_\theta$;
the subscript ``azi'' identifies the azimuthal component.

For the A-series, the cell-centered field saved for the diagnostics is the
arithmetic mean of the two adjacent azimuthal face fields. Spectra and
density--field covariance use this saved field. On the uniform azimuthal
mesh, this operation multiplies a mode's face-field amplitude by
$\cos(\pi n/N_\theta)$, equal to 0.773 and 0.741 for $n=14$ and 15.

At this fixed position and time, $\langle\cdot\rangle_\theta$ denotes the
arithmetic mean over these 64 samples, and the prime in $E'_{{\rm azi},j}$
denotes the departure from that instantaneous azimuthal mean. The local
complex Fourier coefficient is $\widehat E_n(r_i,z_\ell,t)$. Suppressing the
common arguments $(r_i,z_\ell,t)$ within the following equations, the mean,
fluctuation, and Fourier coefficient are computed as
\begin{equation}
  \langle E_{\rm azi}\rangle_\theta
    =\frac{1}{N_\theta}\sum_{j=0}^{N_\theta-1}E_{{\rm azi},j},
  \label{eq:azimuthal-mean}
\end{equation}
\begin{equation}
  E'_{{\rm azi},j}=E_{{\rm azi},j}-\langle E_{\rm azi}\rangle_\theta,
  \label{eq:azimuthal-fluctuation}
\end{equation}
\begin{equation}
  \widehat E_n=\frac{1}{N_\theta}\sum_{j=0}^{N_\theta-1}
     E'_{{\rm azi},j}\exp\!\left(-\mathrm{i}\frac{2\pi nj}{N_\theta}\right).
  \label{eq:azimuthal-dft}
\end{equation}
For $n=1,\ldots,31$, the local one-sided amplitude is
$a_n=2|\widehat E_n|$, with units of $\mathrm{V\,m^{-1}}$. The zero mode is removed by mean subtraction, and the Nyquist mode $n=32$ is excluded from the reported spectra.

Each retained mode is assigned a centerline-reference wavenumber $k_n$.
Using the channel centerline radius $r_c$ and the sector angular width defined
in Eq.~\eqref{eq:sector-definition},
\begin{equation}
  L_{\theta,c}=r_c\Delta\theta_{\rm dom}=6.4~\mathrm{mm},
  \label{eq:reference-length}
\end{equation}
\begin{equation}
  k_n=\frac{2\pi n}{L_{\theta,c}},
  \label{eq:reference-wavenumber}
\end{equation}
\begin{equation}
  \Delta k=\frac{2\pi}{L_{\theta,c}}.
  \label{eq:wavenumber-spacing}
\end{equation}
The spacing $\Delta k=k_{n+1}-k_n$ is approximately
$0.982~\mathrm{rad\,mm^{-1}}$. The horizontal coordinate labeled
$k_{\theta,c}$ in the figures takes these discrete values $k_n$.
At radius $r$, the same sector mode has the local physical wavenumber
\begin{equation}
 k_{\theta,n}(r)=\frac{2\pi n}{r\Delta\theta_{\rm dom}}=\frac{r_c}{r}k_n.
 \label{eq:local-wavenumber}
\end{equation}
A fixed reference-wavenumber band therefore compares the same sector-mode indices,
while their local physical wavelengths vary with radius.
For W0, the azimuthal component is $E_y$; the same mean subtraction and
Fourier transform are applied along its periodic $y$ direction of length
$L_y=6.4~\mathrm{mm}$.

To compare spatial distributions in A1--A4, the active domain is divided
into four regions. The channel has
$r_{\mathrm{ch,in}}\le r\le r_{\mathrm{ch,out}}$ and $z<10~\mathrm{mm}$;
the plume spans the full active radial extent at $z\ge10~\mathrm{mm}$.
Splitting each at $r_c$ gives the channel-inner, channel-outer, plume-inner,
and plume-outer regions, with inner and outer cells selected by $r_i\le r_c$
and $r_i>r_c$, respectively. Solid and inactive cells are excluded.
Let $R$ denote the set of radial--axial cells in any one of these regions.
For each retained mode $n$, its regional mean amplitude is
\begin{equation}
  A_{R,n}(t)=\frac{\sum_{(i,\ell)\in R}V_{i\ell}\,a_n(r_i,z_\ell,t)}
                       {\sum_{(i,\ell)\in R}V_{i\ell}},
  \label{eq:region-spectrum}
\end{equation}
\begin{equation}
  V_{i\ell}=\frac12(r_{i,+}^2-r_{i,-}^2)
                    \Delta\theta_{\rm dom}\,\Delta z_\ell.
  \label{eq:cell-sector-volume}
\end{equation}
Here $V_{i\ell}$ is the sector volume represented by cell $(i,\ell)$,
$r_{i,-}$ and $r_{i,+}$ are its radial faces, and $\Delta z_\ell$ is its
axial width. The points $(k_n,A_{R,n}(t))$ form the regional spectrum,
equivalently $A_R(k_n,t)\equiv A_{R,n}(t)$. Taking $R$ as the complete
active domain defines $A_{{\rm global},n}(t)$, denoted by $A_{\rm global}$
in the figures.

All global spectra in Fig.~\ref{fig:spectral-comparison} use the
physical-volume mean of local amplitudes over active plasma cells.
For W0, the uniform cell volume reduces this to the arithmetic mean over
active radial--axial cells. The cylindrical cases use the sector-volume
weights in Eq.~\eqref{eq:region-spectrum}. Solid and inactive cells are
excluded from every panel.

\paragraph*{Time window and band metrics}
For A1--A4, field snapshots are saved every 1000 electron steps, corresponding to
$0.005~\mu\mathrm{s}$. Quantitative regional spectra use all 401 snapshots
in the common $13$--$15~\mu\mathrm{s}$ window. The time index is
$m=0,\ldots,400$, with $t_m=(13+0.005m)~\mu\mathrm{s}$.
First, each snapshot yields the instantaneous regional amplitudes
$A_{R,n}(t_m)$ from Eq.~\eqref{eq:region-spectrum}; these amplitudes are
then averaged in time by trapezoidal quadrature:
\begin{equation}
\overline A_{R,n}=
  \frac{\sum_{m=0}^{400}\omega_m A_{R,n}(t_m)}
       {\sum_{m=0}^{400}\omega_m}.
  \label{eq:time-averaged-spectrum}
\end{equation}
The endpoint weights are $\omega_0=\omega_{400}=1/2$, and
$\omega_m=1$ for $m=1,\ldots,399$, giving
$\sum_{m=0}^{400}\omega_m=400$. Thus $\overline A_{R,n}$ is the
time-averaged amplitude of mode $n$ in region $R$. Temporal variability
is evaluated in four consecutive $0.5~\mu\mathrm{s}$ blocks, each containing
101 snapshots and using the same endpoint-weighting rule.

For a reference-wavenumber band $\mathcal B$, the condition
$n:\,k_n\in\mathcal B$ selects the retained modes whose $k_n$ lie in that
band. The band amplitude $S_R$, amplitude-weighted mean wavenumber
$\bar k_R$, peak-mode index $n_{\mathrm{peak},R}$, and peak wavenumber
$k_{\mathrm{peak},R}$ are defined by
\begin{equation}
  S_R(\mathcal B)=\sum_{n:\,k_n\in\mathcal B}\overline A_{R,n}\,\Delta k,
  \label{eq:spectral-metrics}
\end{equation}
\begin{equation}
  \bar k_R=\frac{\sum_{n:\,k_n\in\mathcal B}
                      k_n\overline A_{R,n}\,\Delta k}{S_R(\mathcal B)},
  \label{eq:mean-wavenumber}
\end{equation}
\begin{equation}
  n_{\mathrm{peak},R}=\underset{n:\,k_n\in\mathcal B}{\mathrm{arg\,max}}
                       \,\overline A_{R,n},
  \label{eq:peak-mode}
\end{equation}
\begin{equation}
  k_{\mathrm{peak},R}=k_{n_{\mathrm{peak},R}}.
  \label{eq:peak-wavenumber}
\end{equation}
The operator $\mathrm{arg\,max}$ returns the mode index with the largest
time-averaged amplitude within the selected band. Both $\bar k_R$ and
$k_{\mathrm{peak},R}$ have units of $\mathrm{rad\,mm^{-1}}$;
$S_R$ has units of $(\mathrm{V\,m^{-1}})(\mathrm{rad\,mm^{-1}})$.
Two fixed bands, $10$--$15~\mathrm{rad\,mm^{-1}}$ ($n=11$--15) and
$3$--$8~\mathrm{rad\,mm^{-1}}$ ($n=4$--8), are used to compare radial
amplitude distributions at different azimuthal scales. The complete
31-mode spectrum is also retained. Channel peak and mean-wavenumber
metrics use $8\le k_n\le18~\mathrm{rad\,mm^{-1}}$ and the spectrum
$\overline A_{{\rm ch},n}=(\overline A_{{\rm ch,in},n}
+\overline A_{{\rm ch,out},n})/2$, which assigns equal weight to the
volume-normalized inner and outer channel spectra. The subscripts
``ch'' and ``pl'' denote channel and plume, and ``in'' and ``out'' denote
their inner and outer parts. Their radial asymmetry is measured by
\begin{equation}
  R_{\mathrm{out/in,ch}}
    =\frac{S_{\mathrm{ch,out}}(10\le k_n\le15)}
            {S_{\mathrm{ch,in}}(10\le k_n\le15)},
  \label{eq:outer-inner-ratios}
\end{equation}
\begin{equation}
  R^{\mathrm{low}}_{\mathrm{out/in,pl}}
    =\frac{S_{\mathrm{pl,out}}(3\le k_n\le8)}
            {S_{\mathrm{pl,in}}(3\le k_n\le8)},
  \label{eq:plume-low-ratio}
\end{equation}
\begin{equation}
  R^{\mathrm{high}}_{\mathrm{out/in,pl}}
    =\frac{S_{\mathrm{pl,out}}(10\le k_n\le15)}
            {S_{\mathrm{pl,in}}(10\le k_n\le15)}.
  \label{eq:plume-high-ratio}
\end{equation}
The two plume bands are reported separately; a ratio close to unity denotes similar
volume-averaged regional amplitudes.

\paragraph*{Density--field covariance}
The density--field covariance connects the electric-field fluctuations to
their correlated axial drift-flux contribution. For the prescribed radial
magnetic field,
\begin{equation}
  \mathbf{v}_E=\frac{\mathbf{E}\times\mathbf{B}}{B^2},
  \label{eq:exb-velocity}
\end{equation}
\begin{equation}
  v_{E,z}=-\frac{E_\theta}{B_r}.
  \label{eq:exb-axial-velocity}
\end{equation}
For the imposed, time-independent $B_r$, which is uniform in $\theta$,
the fluctuating electron particle flux associated with this drift is
\begin{equation}
  \Gamma_{e,z}^{E\times B,\rm fluct}
  =\langle\delta n_e\,\delta v_{E,z}\rangle_{\theta,t}
    =-\frac{C_{nE}}{B_r},
  \label{eq:fluctuating-axial-flux}
\end{equation}
\begin{equation}
  C_{nE}=\langle\delta n_e\,\delta E_\theta\rangle_{\theta,t},
  \label{eq:density-field-covariance}
\end{equation}
where $\langle\cdot\rangle_{\theta,t}$ is the joint azimuthal--time average
over the common window and $\delta f=f-\langle f\rangle_{\theta,t}$. This joint average combines the arithmetic azimuthal mean with the trapezoidal time weights used for the spectra.
Matched density and field snapshots give
$C_{nE}=\langle n_eE_\theta\rangle_{\theta,t}
-\langle n_e\rangle_{\theta,t}\langle E_\theta\rangle_{\theta,t}$,
with units of $\mathrm{V\,m^{-4}}$.
The covariance measures how density variations align with the axial
$E\times B$ drift: a preferred association of higher density with
anode-directed drift produces an anode-directed mean flux contribution.
For $B_r>0$, this contribution corresponds to $C_{nE}>0$.
To compare the radial distribution of this density--field coupling, define
the regional mean covariance magnitude and its outer/inner ratios as
\begin{equation}
  Q_R=\frac{\int_R|C_{nE}|\,dV}{\int_RdV},
  \label{eq:regional-covariance}
\end{equation}
\begin{equation}
  R_{C,\rm ch}=\frac{Q_{\rm ch,out}}{Q_{\rm ch,in}},
  \label{eq:covariance-ratio}
\end{equation}
\begin{equation}
  R_{C,\rm pl}=\frac{Q_{\rm pl,out}}{Q_{\rm pl,in}}.
  \label{eq:plume-covariance-ratio}
\end{equation}
The same four regions are used as for the spectra. $Q_R$ measures the mean
magnitude of density--field coupling; its flux contribution is obtained from
the signed $-C_{nE}/B_r$ using the local magnetic field. The regional ratios
therefore characterize the covariance distribution. The signed integral
$\int_R C_{nE}\,dV$ is also evaluated to quantify sign cancellation.

\paragraph*{Numerical resolution}
All cylindrical cases use $N_\theta=64$ azimuthal cells, so the cell width at the channel
centerline is $\Delta s_{\theta,c}=L_{\theta,c}/N_\theta=0.1~\mathrm{mm}$. The common-window
channel peaks at $n=14$--15 have centerline wavelengths of $0.427$--$0.457~\mathrm{mm}$,
or about $4.3$--$4.6$ azimuthal cells per wavelength. These fixed-resolution comparisons characterize the spectral and radial patterns at this sampling scale. Finer azimuthal-grid comparisons would quantify the resolution dependence of spectral amplitudes and outer/inner ratios.

\section{Simulation results}
\label{sec:results}

The comparison begins with global evolution and matched-time fields,
then examines spectral and radial structure and density--field covariance.
The A1--A4 late-time diagnostics use $13$--$15~\mu\mathrm{s}$.

\subsection{Global evolution and instantaneous spatial structure}
\begin{figure}[!htbp]
\centering
  \includegraphics[width=\columnwidth]{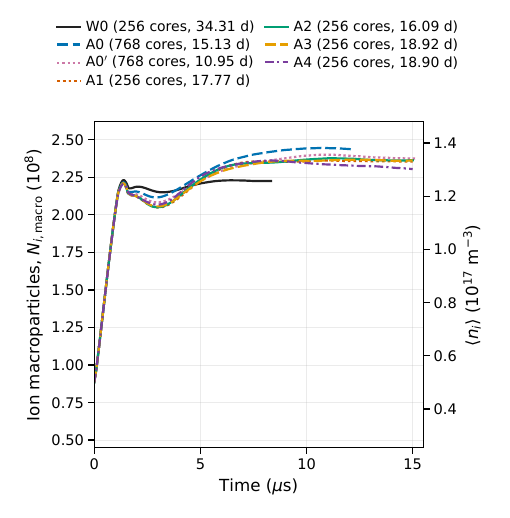}
  \caption{Evolution of total ion macroparticle number (left axis) and corresponding
domain-averaged ion density (right axis) for W0, A0, A0$'$, and A1--A4.
The legend gives the case, physical CPU cores, and cumulative loop time
in days for the accepted trajectories listed in Table~\ref{tab:production-cost}.
Timing definitions are given in Sec.~\ref{sec:production-efficiency}.
W0 is plotted from its first post-loading diagnostic sample.}
  \label{fig:global-evolution}
\end{figure}

Figure~\ref{fig:global-evolution} shows a common initial rise, overshoot,
and relaxation, followed by slower population evolution.

Over the common $7$--$8~\mu\mathrm{s}$ interval, the mean ion macroparticle
numbers are $2.22551\times10^8$, $2.39970\times10^8$,
$2.34318\times10^8$, and $2.35302\times10^8$ in W0, A0, A0$'$, and A1,
respectively. A0, A0$'$, and A1 exceed W0 by
$7.83\%$, $5.29\%$, and $5.73\%$. Thus the near-Cartesian cases follow
the same broad transient sequence but retain finite population differences.

Related differences between PIC realizations and implementations have
also been reported in Hall-thruster benchmarks. Charoy et al. repeated one
axial--azimuthal case three times with identical input parameters and found
order-$5\%$ differences in time-averaged ion-density and axial-field
profiles; fixing the random seed recovered overlapping current
histories\cite{Charoy2019PSST}. In the radial--azimuthal benchmark of
Villafana et al., averaged ion-density and total electron-temperature
profiles remained within $\pm2.5\%$ of the multi-code mean, while using
common initial particle positions substantially reduced the transient
differences\cite{Villafana2021PSST}. Together, the common-initialization
and repeated-seed tests show that PIC results are sensitive to particle
sampling, including initialization, and can exhibit statistical variability
even with the same nominal physical inputs. Such variability is one
possible source of differences between simulation results.

The population differences observed here may also reflect the numerical settings.
W0 and the A-series differ in particle--field coupling and cathode-injection timing,
as detailed in Sec.~II B.
A0$'$ introduces ten-step ion supercycling relative to A0. A1 retains this
ion-update scheme while introducing a non-uniform radial--axial mesh and
MPI load balancing.
These differences may affect particle residence and boundary losses through
changes in the self-consistent fields and particle trajectories.

Over $13$--$15~\mu\mathrm{s}$, the mean populations in A2, A3, and A4
differ from A1 by $+0.29\%$, $+0.24\%$, and $-1.74\%$, respectively.
These curvature cases therefore have similar bulk populations.

\begin{figure*}[!htbp]
  \centering

  \includegraphics[width=\textwidth]{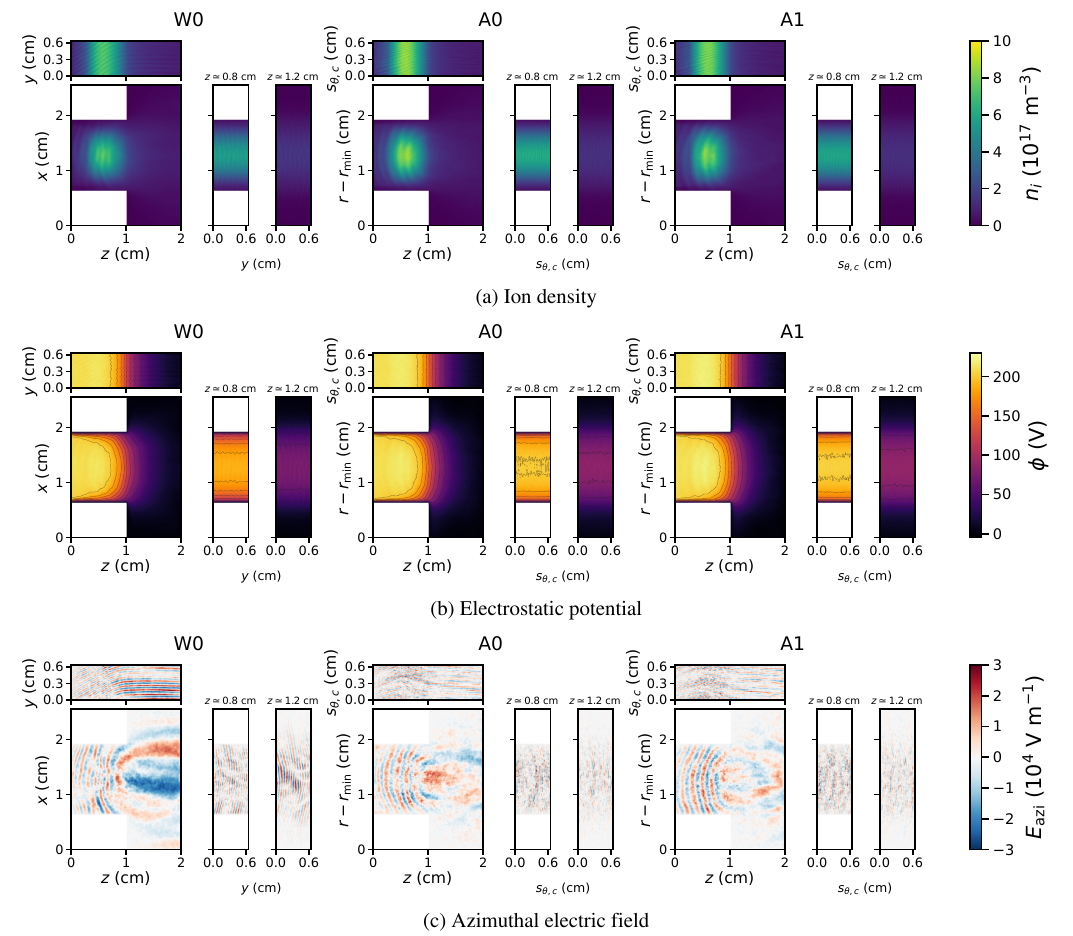}
  \caption{Instantaneous fields at $8~\mu\mathrm{s}$ for W0, A0, and A1.
Rows show ion density, potential, and azimuthal electric field
($E_y$ for W0, $E_\theta$ for the cylindrical cases).
Potential contours are spaced by 20 V.\ 
Each row uses a shared color scale across cases.\ 
Cuts are taken at the radial and azimuthal midpoints and near
$z=0.8$ and $1.2~\mathrm{cm}$. Cylindrical positions use $r-r_{\min}$
and centerline arc length $s_{\theta,c}$; W0 uses $x,y$.}
  \label{fig:instantaneous-structures}
\end{figure*}

Figure~\ref{fig:instantaneous-structures} compares W0, A0, and A1 at
$8~\mu\mathrm{s}$. All three have a channel-centered ion-density maximum,
a lower-density plume, and a potential that decreases toward the outlet.
W0 has longer azimuthal-field wavelengths than A0 and A1. To check the W0--A0
difference at the same location, their $8~\mu\mathrm{s}$ fields are sampled
at $x=12.75~\mathrm{mm}$ for W0 and $r-r_{\min}=12.75~\mathrm{mm}$ for A0,
near the channel centerline. At each
axial position inside the channel ($z<10~\mathrm{mm}$), the 64 azimuthal
samples over $6.4~\mathrm{mm}$ are transformed and their amplitudes are then
averaged along $z$. These channel-cut spectra peak at $n=8$ for W0 and
$n=13$ for A0, corresponding to wavelengths of 0.800 and
$0.492~\mathrm{mm}$, respectively.

\subsection{Spectral evolution and radial partition}
\begin{figure}[!htbp]
  \centering
  \includegraphics[width=\columnwidth]{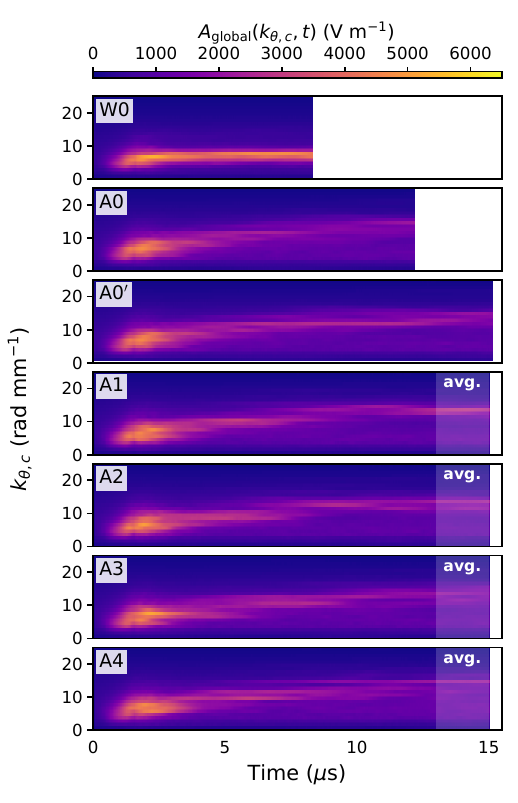}
  \caption{Evolution of one-sided azimuthal electric-field amplitude spectra in all seven cases.
The centerline-reference azimuthal length is $6.4~\mathrm{mm}$.
For A1--A4, the spectra shown at $t<13~\mu\mathrm{s}$ use one in every ten saved
field snapshots, giving a display interval of $0.05~\mu\mathrm{s}$; all snapshots
are used over $13$--$15~\mu\mathrm{s}$, at $0.005~\mu\mathrm{s}$ intervals.
The W0, A0, and A0$'$ panels use their available spectra at
$0.005~\mu\mathrm{s}$ intervals. Shading marks the common window used for
quantitative statistics; blank intervals indicate unavailable outputs.
All panels use physical-volume mean amplitudes over active plasma cells,
as defined in Sec.~\ref{sec:diagnostics}.}
  \label{fig:spectral-comparison}
\end{figure}

\begin{figure*}[!htbp]
  \centering
  \includegraphics[width=0.98\textwidth]{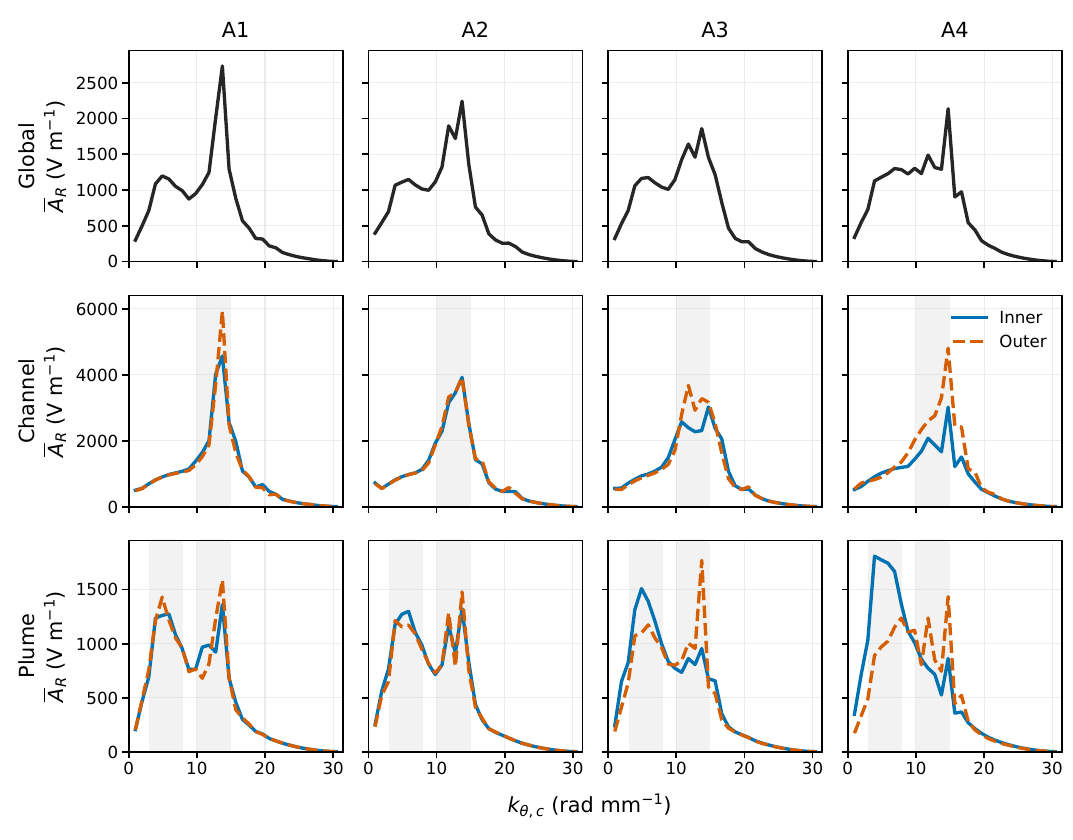}
  \caption{Azimuthal electric-field amplitudes averaged over $13$--$15~\mu\mathrm{s}$.
Top: active-volume global spectrum. Middle and bottom: inner and outer
channel and plume spectra. Shading denotes the channel band
$10$--$15~\mathrm{rad\,mm^{-1}}$ and the plume bands
$3$--$8$ and $10$--$15~\mathrm{rad\,mm^{-1}}$.}
  \label{fig:radial-spectra}
\end{figure*}

Figure~\ref{fig:spectral-comparison} shows the evolution of the W0--A0
spectral separation. Both global spectra peak at $n=7$ at $2~\mu\mathrm{s}$.
At $8~\mu\mathrm{s}$, W0 peaks at $n=8$ and A0 at $n=12$, corresponding
to $7.854$ and $11.781~\mathrm{rad\,mm^{-1}}$, or wavelengths of $0.800$
and $0.533~\mathrm{mm}$. Averaging over $7$--$8~\mu\mathrm{s}$ retains
these peaks; A0$'$ also peaks at $n=12$, while A1 peaks at $n=11$.
The scale separation thus appears in both matched spatial cuts and
time-averaged spectra. The cross-code comparison therefore shows similar broad population evolution together with a persistent difference in the dominant spectral scale. The curvature-dependent radial comparisons below use the common A1--A4 implementation.

Differences between W0 and the A-series in charge representation and filtering,
Poisson discretization, field staggering, and field gathering may alter the
finite-grid response to charge perturbations at different wavenumbers and the
competition between modes, thereby affecting the dominant wavelength~\cite{Huang2016SpectralFidelity}.

Within the controlled A1--A4 curvature scan, all four independently initialized cases
develop appreciable azimuthal fluctuations during the early population rise. Their
spectral weight subsequently shifts toward higher centerline-reference wavenumbers.
The common-window channel spectra peak at sector index $n=14$ in A1 and A2 and
$n=15$ in A3 and A4. These correspond to $13.744$ and
$14.726~\mathrm{rad\,mm^{-1}}$, respectively, with the resolved spacing
$\Delta k_{\theta,c}=2\pi/L_{\theta,c}=0.982~\mathrm{rad\,mm^{-1}}$.
The peak difference is therefore one resolved Fourier bin.

\Needspace{3\baselineskip}

The channel centroids in the $8$--$18~\mathrm{rad\,mm^{-1}}$ band span
$12.904$--$13.310~\mathrm{rad\,mm^{-1}}$
(Table~\ref{tab:curvature-estimate}). These band-limited centroids differ from the
full-spectrum centroids, which are $12.227$, $11.916$, $12.164$, and
$12.107~\mathrm{rad\,mm^{-1}}$ in A1--A4. The lower values reflect the weight of
the low-wavenumber content. Both measures indicate a broadly similar spectral scale across the curvature scan.

\begin{table}[!htbp]
\centering
\caption{Diagnostics over $13$--$15~\mu\mathrm{s}$.
Channel peaks and centroids use the equal-weight inner--outer spectrum;
centroids use $8$--$18~\mathrm{rad\,mm^{-1}}$.
Spectral ratios use the indicated bands, and $R_C$ uses regional mean covariance
magnitudes.}
  \label{tab:curvature-estimate}
\footnotesize
\setlength{\tabcolsep}{2pt}
\renewcommand{\arraystretch}{1.12}
\begin{tabular}{lrrrr}
\toprule
Metric & A1 & A2 & A3 & A4 \\
\midrule
$k_{\mathrm{peak,ch}}$ ($\mathrm{rad\,mm^{-1}}$) & 13.744 & 13.744 & 14.726 & 14.726 \\
$\bar{k}_{\mathrm{ch}}$, $8$--$18$ ($\mathrm{rad\,mm^{-1}}$) & 13.271 & 12.904 & 13.147 & 13.310 \\
Channel $10$--$15$: $R_{\mathrm{out/in,ch}}$ & 1.049 & 1.018 & 1.260 & 1.532 \\
Plume $3$--$8$: $R^{\mathrm{low}}_{\mathrm{out/in,pl}}$ & 1.011 & 0.955 & 0.835 & 0.630 \\
Plume $10$--$15$: $R^{\mathrm{high}}_{\mathrm{out/in,pl}}$ & 1.016 & 1.015 & 1.283 & 1.358 \\
Channel covariance: $R_{C,\mathrm{ch}}$ & 1.098 & 1.030 & 1.251 & 2.714 \\
Plume covariance: $R_{C,\mathrm{pl}}$ & 1.068 & 0.942 & 0.787 & 0.578 \\
\bottomrule
\end{tabular}
\end{table}

\begin{figure*}[!htbp]
  \centering
  \includegraphics[width=0.98\textwidth]{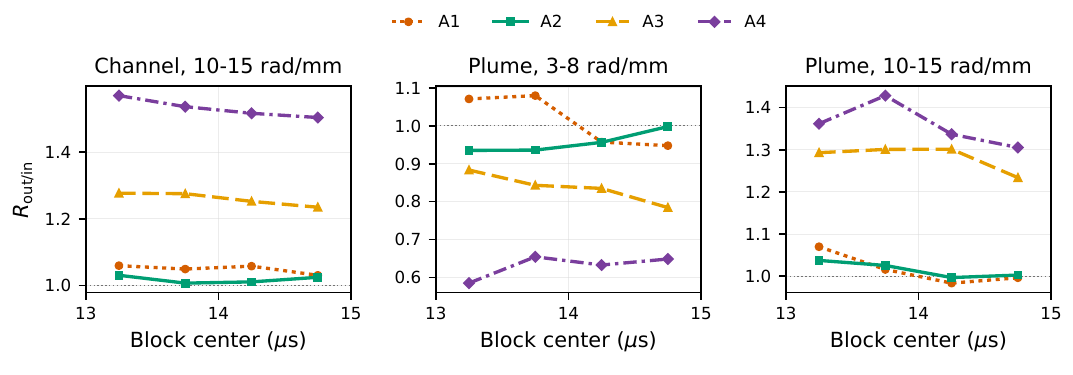}
  \caption{Spectral outer/inner amplitude ratios in four consecutive
$0.5~\mu\mathrm{s}$ blocks over $13$--$15~\mu\mathrm{s}$.
Unity denotes equal regional mean amplitudes.}
  \label{fig:block-evolution}
\end{figure*}

The regional spectra (Fig.~\ref{fig:radial-spectra}) show a stronger contrast than
the peaks. Channel band amplitudes are nearly balanced in A1--A2 but outer weighted
in A3--A4 (ratios 1.260 and 1.532). A2 is closest to radial balance. In the plume, the low-wavenumber ratios fall to
0.835 and 0.630 in A3 and A4, whereas the high-wavenumber ratios reach 1.283 and
1.358. Broadband averaging would hide these opposing radial preferences.

Figure~\ref{fig:block-evolution} shows that the curvature-associated radial
partition persists in every half-microsecond block. A1--A2 remain close to
unity, whereas A3--A4 consistently have ratios above unity in the channel
and high-wavenumber plume band, and below unity in the low-wavenumber plume
band. A4 departs further from unity than A3 in all three diagnostics in
every block. The channel ratios decrease from 1.277 to 1.235 in A3 and
from 1.570 to 1.505 in A4 without reversing their outer preference.
Thus stronger curvature is associated with a persistent, band-dependent
radial asymmetry over $13$--$15~\mu\mathrm{s}$.

The channel ratio also depends on which spectral content is integrated. Expanding the
original $n=11$--15 band to $n=10$--16 gives ratios of
1.015, 1.013, 1.182, and 1.563 for A1--A4. For A3, contracting the band to
$n=12$--14 gives 1.416, while integrating the wider
$8$--$18~\mathrm{rad\,mm^{-1}}$ interval gives 1.104. A4 remains distinctly
outer weighted under these alternatives, but the numerical magnitude of the A3
asymmetry is appreciably band dependent. These checks show the band dependence of the radial redistribution.

\Needspace{3\baselineskip}
\subsection{Density--field covariance}
\begin{figure}[!htbp]
  \centering
  \includegraphics[width=\columnwidth]{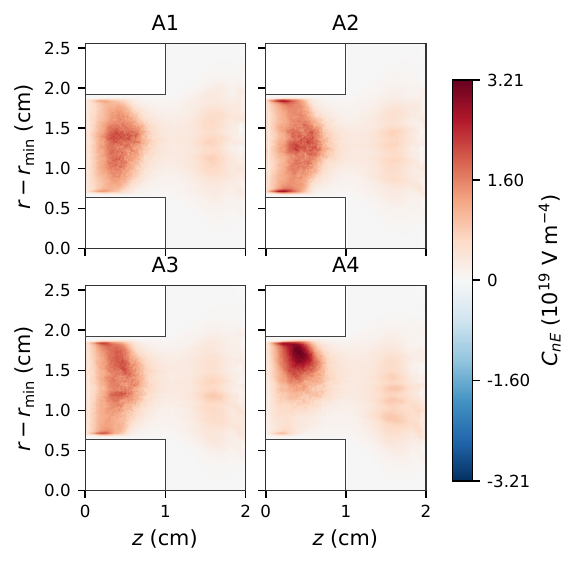}
  \caption{Density--field covariance
$C_{nE}=\langle\delta n_e\,\delta E_\theta\rangle_{\theta,t}$
averaged over $13$--$15~\mu\mathrm{s}$.
All panels share the symmetric color scale; white regions are solid or inactive.}
  \label{fig:density-field-covariance}
\end{figure}

\begin{figure}[!htbp]
  \centering
  \includegraphics[width=0.75\columnwidth,trim=175bp 170bp 140bp 0bp,clip]{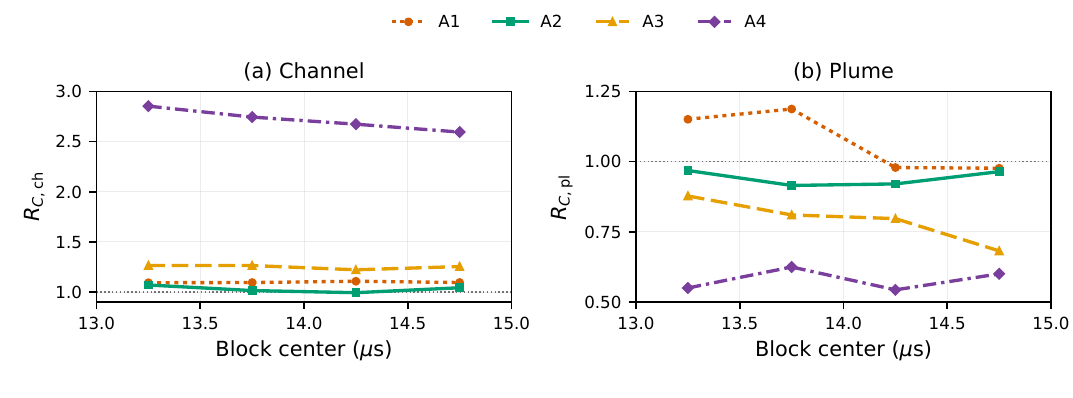}\par
  \includegraphics[width=0.92\columnwidth,trim=0bp 0bp 257.4bp 23bp,clip]{figures/covariance_blocks_20261029/covariance_ratio_blocks.pdf}\par
  \includegraphics[width=0.92\columnwidth,trim=257.4bp 0bp 0bp 23bp,clip]{figures/covariance_blocks_20261029/covariance_ratio_blocks.pdf}
  \caption{Covariance-magnitude outer/inner ratios for A1--A4 in the
(a) channel and (b) plume, evaluated in four consecutive $0.5~\mu\mathrm{s}$
blocks over $13$--$15~\mu\mathrm{s}$. Each block uses its own joint
azimuthal--time means and physical-volume regional averages of $|C_{nE}|$.
Markers denote block centers; connecting lines guide the eye. Unity denotes
equal regional mean magnitudes. The panels use different vertical ranges.}
  \label{fig:covariance-block-evolution}
\end{figure}

The covariance in Fig.~\ref{fig:density-field-covariance} is predominantly positive,
corresponding to an electron $E\times B$ drift contribution toward the anode for
the prescribed positive $B_r$. The channel covariance-magnitude ratios are
1.098, 1.030, 1.251, and 2.714 in A1--A4, while the plume ratios are
1.068, 0.942, 0.787, and 0.578 (Table~\ref{tab:curvature-estimate}).
The strongest-curvature case therefore combines a strongly outer-weighted channel
covariance with an inner-weighted plume covariance. A1 and A2 remain much closer
to radial balance than A4 in this channel diagnostic.

Evaluating the signed integral and the integral of the magnitude over each of the
four regions gives
$|\int_R C_{nE}\,dV|/\int_R|C_{nE}|\,dV\ge0.9974$ for all cases.
Regional sign cancellation is consequently small.

The covariance identifies where density and electric-field fluctuations
combine to contribute to axial drift transport. In A4, this coupling is
concentrated in the outer channel and inner plume.

Figure~\ref{fig:covariance-block-evolution} shows a persistent late-time radial
preference. The channel ratios remain near unity in A1--A2, but stay within
1.222--1.266 in A3 and 2.595--2.853 in A4. The corresponding plume ratios
remain below unity throughout, within 0.682--0.878 and 0.544--0.625.
Consequently, the stronger-curvature cases maintain outer-channel and
inner-plume covariance preferences in every block, with a larger departure
from unity in A4 than in A3. Together with Fig.~\ref{fig:block-evolution},
this establishes persistent curvature-associated radial partitioning
over $13$--$15~\mu\mathrm{s}$ in both field amplitudes and density--field
coupling.

\Needspace{3\baselineskip}
\subsection{Computational efficiency and accepted-trajectory cost}
\label{sec:production-efficiency}

\begin{figure*}[!htbp]
\centering
\includegraphics[width=\textwidth]{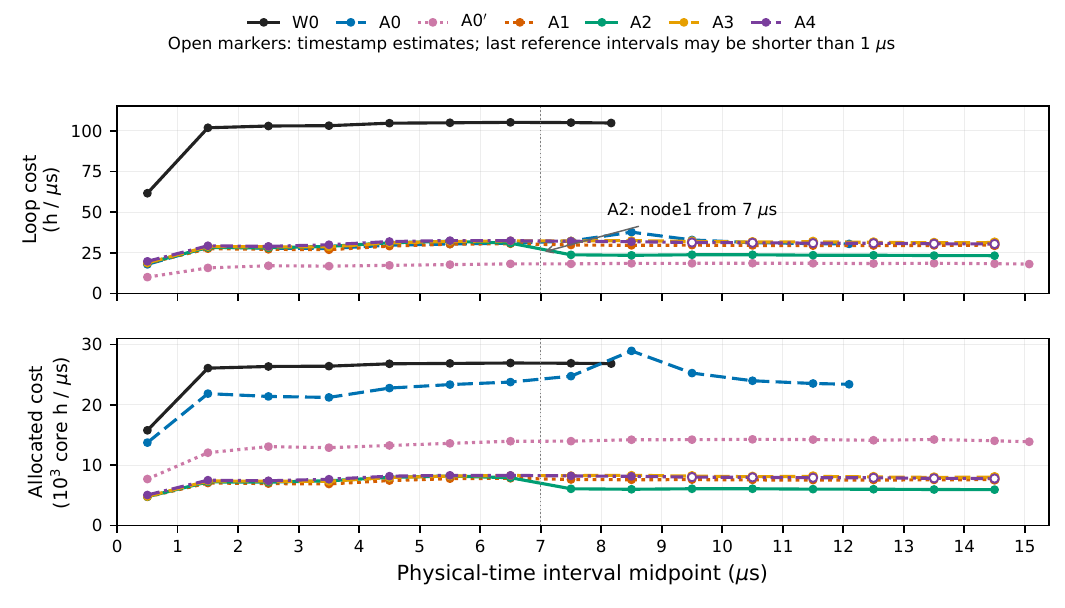}
\caption{Accepted-trajectory cost per physical microsecond in successive intervals.
Top: loop hours; bottom: allocated core-hours. Each value is normalized by its
interval length. Open markers indicate timestamp-recovered intervals.
The dotted line marks A2's migration to EPYC 9965 at $7~\mu\mathrm{s}$.}
  \label{fig:computational-cost-evolution}
\end{figure*}

We compare elapsed loop time and core-hour cost over matched physical-time
intervals. For a loop duration $T_{\rm loop}$ in hours and $P$ physical CPU
cores, the cost per simulated microsecond is
\begin{equation}
 C_{\rm phys}=\frac{P T_{\rm loop}}{\Delta t_{\rm phys}},
\end{equation}
where $\Delta t_{\rm phys}$ is in microseconds. Costs count each retained
trajectory interval once and include in-loop output, but exclude setup,
restart loading, queueing, migration, and downtime.

W0 uses $64$ MPI ranks and four OpenMP threads ($256$ cores) on EPYC 9754.
A0 and A0$'$ use $384$ ranks and two threads ($768$ cores) on two EPYC 9965
nodes. A1--A4 use $128$ ranks and two threads ($256$ cores). A1, A3, and A4
use EPYC 9754; A1 migrates between nodes at $5.5~\mu\mathrm{s}$.
A2 switches from EPYC 9754 to EPYC 9965 at $7~\mu\mathrm{s}$.

\begin{table}[!htbp]
\centering
\caption{Accepted-trajectory costs through each case's final physical
time. $P$ is the number of bound physical cores. A1, A3, and A4 include
timestamp-recovered segments. The different final times are retained explicitly.}
  \label{tab:production-cost}
\normalsize\renewcommand{\arraystretch}{1.20}\setlength{\tabcolsep}{3pt}
\begin{tabular}{lrrrr}
\toprule
Case & $P$ & Final $t$ ($\mu$s) & Loop (h) & Core h/$10^3$ \\
\midrule
W0 & 256 & 8.3285 & 823.4 & 210.80 \\
A0 & 768 & 12.1925 & 363.2 & 278.94 \\
A0$'$ & 768 & 15.1455 & 262.8 & 201.83 \\
A1 & 256 & 15 & 426.5 & 109.19 \\
A2 & 256 & 15 & 386.1 & 98.84 \\
A3 & 256 & 15 & 454.0 & 116.23 \\
A4 & 256 & 15 & 453.7 & 116.14 \\
\bottomrule
\end{tabular}
\end{table}

Table~\ref{tab:production-cost} lists cumulative costs, while
Table~\ref{tab:matched-production-cost} separates elapsed-time gains from
core-hour savings. Over $0$--$8~\mu\mathrm{s}$, W0 takes 3.33--3.59 times
the elapsed time and core-hours of A1--A4 at the same core count. The larger
elapsed-time gains of A0 and A0$'$ partly reflect their threefold core count.
Over $13$--$15~\mu\mathrm{s}$, A1--A4 use 43--58\% fewer core-hours than
A0$'$ but take longer elapsed times.

\begin{table}[!htp]
\centering
\caption{Matched-window particle counts and production costs.
The upper block uses W0 as reference over $0$--$8~\mu\mathrm{s}$;
the lower block uses A0$'$ over $13$--$15~\mu\mathrm{s}$.
Mean total particle counts use uniform 100-step sampling.
Ratios equal reference cost divided by case cost.}
  \label{tab:matched-production-cost}
\footnotesize\setlength{\tabcolsep}{2.5pt}
\begin{tabular}{lrrrrrr}
\toprule
Case & \shortstack{$\overline{N_e+N_i}$\\$/10^8$} & \shortstack{Loop time\\(h)} & h/$\mu$s & \shortstack{Core h/$\mu$s\\$/10^3$} & \shortstack{Time\\ratio} & \shortstack{Core h\\ratio} \\
\midrule
\multicolumn{7}{c}{\shortstack{$0$--$8~\mu\mathrm{s}$; reference: W0;\\all counters directly recorded}} \\
W0 & 4.187 & 789.0 & 98.63 & 25.249 & 1.00 & 1.00 \\
A0 & 4.272 & 225.0 & 28.12 & 21.596 & 3.51 & 1.17 \\
A0$'$ & 4.208 & 130.9 & 16.36 & 12.562 & 6.03 & 2.01 \\
A1 & 4.197 & 219.7 & 27.46 & 7.031 & 3.59 & 3.59 \\
A2 & 4.201 & 221.6 & 27.71 & 7.093 & 3.56 & 3.56 \\
A3 & 4.189 & 232.4 & 29.05 & 7.437 & 3.39 & 3.39 \\
A4 & 4.217 & 236.7 & 29.59 & 7.576 & 3.33 & 3.33 \\
\midrule\multicolumn{7}{c}{$13$--$15~\mu\mathrm{s}$; reference: A0$\prime$} \\
A0$'$ & 4.729 & 36.8 & 18.41 & 14.136 & 1.00 & 1.00 \\
A1 & 4.683 & 59.2 & 29.61 & 7.580 & 0.62 & 1.86 \\
A2 & 4.697 & 46.5 & 23.24 & 5.950 & 0.79 & 2.38 \\
A3 & 4.694 & 62.4 & 31.20 & 7.988 & 0.59 & 1.77 \\
A4 & 4.602 & 60.9 & 30.46 & 7.799 & 0.60 & 1.81 \\
\bottomrule
\end{tabular}
\end{table}

\par\noindent\begin{minipage}{\columnwidth}
\setlength{\parindent}{10pt}\indent
The A0 cost peak near $8~\mu\mathrm{s}$ in
Fig.~\ref{fig:computational-cost-evolution} is caused by a defect in the
multi-node restart particle assignment at $7.9~\mu\mathrm{s}$. Code
inspection and targeted selector tests reproduced the imbalance: two MPI
ranks on different nodes represent the same physical region, but one
receives most particles and the other very few, slowing synchronized
advancement. The defect changes execution ownership while preserving the
restored particle positions, velocities, and global count. It is therefore
a parallel load-distribution error, not an error in restoring the particle
physical state. The imbalance eventually relaxes; the extra runtime is
included in the reported costs.
\end{minipage}\par

\Needspace{12\baselineskip}
\section{Discussion and Conclusions}
\label{sec:discussion}
\label{sec:conclusion}

Across the present radius scan, the bulk plasma populations and dominant channel wavenumbers on the common centerline-reference scale remain similar. The weak variation in bulk populations is qualitatively consistent with the similar azimuthally averaged radial density profiles found at a selected axial location in the three-dimensional cylindrical--Cartesian comparison of Taccogna and Minelli \cite{taccognaThreedimensionalParticleincellModel2018}. The present simulations extend the curvature comparison to the spatial organization of fluctuations throughout a channel--plume domain, with radial, azimuthal, and axial dynamics evolving self-consistently as the radius varies at fixed channel width and imposed magnetic-field profile.

The principal additional finding is the radial redistribution of fluctuations. At $r_{\min}=10~\mathrm{cm}$ and $2~\mathrm{cm}$, the outer-channel electric-field amplitudes in the $10$--$15~\mathrm{rad\,mm^{-1}}$ reference-wavenumber band exceed the inner-channel values by $26\%$ and $53\%$, respectively. In the plume, the low-wavenumber band ($3$--$8~\mathrm{rad\,mm^{-1}}$) favors the inner region, whereas the high-wavenumber band ($10$--$15~\mathrm{rad\,mm^{-1}}$) favors the outer region; their outer-to-inner ratios at $r_{\min}=2~\mathrm{cm}$ are 0.63 and 1.36. Thus, the curvature dependence varies with both axial region and spectral band, and broadband or spatially averaged measures can conceal opposing radial preferences.

Density--electric-field covariance further identifies where fluctuations contribute to the correlated axial $E\times B$ drift. At $r_{\min}=2~\mathrm{cm}$, its outer-to-inner magnitude ratios are 2.71 in the channel and 0.58 in the plume. The local electron particle-flux contribution is $-C_{nE}/B_r$; the predominantly positive covariance therefore corresponds to an anode-directed contribution for the prescribed positive radial field. These radial preferences persist throughout $13$--$15~\mu\mathrm{s}$, when the bulk plasma populations are approximately steady. The covariance complements the field spectra by identifying where density and electric-field fluctuations jointly contribute to axial drift, linking the geometric redistribution of wave amplitudes to the spatial organization of transport-related coupling.

Consequently, agreement in global populations or dominant reference wavelengths alone does not establish agreement in local fluctuation and transport-related structure. Detailed comparisons with spatially resolved experiments should account for the radial and axial distributions of both field fluctuations and density--field coupling. For the $1.28~\mathrm{cm}$ channel width studied, the changes at $r_{\min}=10~\mathrm{cm}$ and $2~\mathrm{cm}$, corresponding to $\chi\approx0.057$ and 0.195, support retaining cylindrical geometry when evaluating these local quantities, even with an unchanged imposed magnetic-field profile. These tested configurations provide quantitative reference points for geometry selection; a general transition radius requires additional radius and channel-width scans.

The present prescribed-source, collisionless model uses absorbing boundaries and a radius-independent radial magnetic field with nonzero divergence. Future work will extend the comparison to divergence-free fields, self-consistent ionization, collisions, secondary-electron emission, and dielectric-wall charging to assess how these spatial trends carry over to experimental thrusters.

\begin{acknowledgments}
The authors acknowledge support from the National Natural Science Foundation of China
under Grant No.~52472403.
\end{acknowledgments}

\section*{Data Availability}

The data that supports the findings of this study are available from the corresponding author
upon reasonable request.

\renewcommand{\bibsection}{\section*{References}}
\bibliographystyle{unsrt}
\bibliography{references}

\end{document}